\documentclass{elsart1p}
\usepackage{graphics}
\usepackage{graphicx}
\usepackage{epsfig}
\usepackage{amssymb}

\begin{document}

\begin{frontmatter}

\title{Improved constraints on chiral SU(3) dynamics\\ from kaonic hydrogen}


\author[a,b]{Yoichi Ikeda\corauthref{cor}},
\ead{yikeda@riken.jp}
\author[a]{Tetsuo Hyodo} \author[c]{and Wolfram Weise}
\address[a]{Department of Physics, Tokyo Institute of Technology, Meguro 152-8551, Japan}
\address[b]{RIKEN Nishina Center, 2-1, Hirosawa, Wako, Saitama 351-0198, Japan}
\address[c]{Physik-Department, Technische Universit\"at M\"unchen, D-85747 Garching, Germany}
\corauth[cor]{Corresponding author at: Department of Physics, Tokyo Institute of Technology, Meguro 152-8551, Japan}



\begin{abstract}
A new improved study of $K^-$-proton interactions near threshold is
performed using coupled-channels dynamics based on the next-to-leading
order  chiral SU(3) meson-baryon effective Lagrangian. Accurate constraints are now provided by new 
high-precision kaonic hydrogen measurements. Together with threshold branching ratios and
scattering data, these constraints permit an updated analysis of the complex $\bar{K}N$ and $\pi\Sigma$ coupled-channels amplitudes and an improved determination of the $K^-p$ scattering length, including uncertainty estimates. 
\end{abstract}

\end{frontmatter}

%


{\it Introduction}. Within the hierarchy of quark masses in QCD, the strange quark plays an intermediate role between ``light" and ``heavy". Hadronic systems with strange quarks  and, in particular,  antikaon-nucleon interactions close to threshold are therefore suitable testing grounds for investigating the interplay between spontaneous and explicit 
chiral symmetry breaking in low-energy QCD. 

Methods of effective field theory with coupled-channels, based on the chiral SU(3)$_{R}\times$SU(3)$_{L}$ meson-baryon effective Lagrangian, have become a well established framework for dealing with low-energy $\bar{K}N$
interactions~\cite{Kaiser1995} (see also Ref.~\cite{HJ2011} for a recent review). However, previous applications of such approaches, combining information from earlier kaonic hydrogen measurements~\cite{Iwasaki1997, Beer2005} and older $K^- p$ scattering data, were still subject to considerable uncertainties. The theoretical studies~\cite{Borasoy2005, Borasoy2006} gave strong indications for a possible inconsistency between the DEAR $K^-$ hydrogen data~\cite{Beer2005} and the low-energy $K^- p$ elastic scattering cross section. With the recent appearance of 
results from the SIDDHARTA kaonic hydrogen measurements~\cite{Bazzi2011}, a new level of accuracy has now been reached that permits an improved analysis with updated constraints. The present work describes such an analysis including a new determination of the $K^-$-proton scattering length and implications for the coupled $\bar{K}N$ and $\pi$-hyperon amplitudes below $K^- p$ threshold. 

{\it Theoretical framework}. The starting point is the chiral SU(3)$_{R}\times$SU(3)$_{L}$ meson-baryon effective Lagrangian at next-to-leading order (NLO):
\begin{eqnarray} 
{\cal L}_{eff} = {\cal L}_M({\cal U}) + {\cal L}_{MB}^{(1)}({\cal B,U}) +  {\cal L}_{MB}^{(2)}({\cal B,U})~~,
\end{eqnarray}
where ${\cal L}_M({\cal U})$ with ${\cal U} = u^2 = \exp[\textrm{i}\sqrt{2}\,\Phi/f]$ is the non-linear chiral meson Lagrangian incorporating the octet of pseudoscalar Nambu-Goldstone bosons $(\pi, K, \bar{K}, \eta)$ in the standard $3\times 3$ matrix representation $\Phi$. At this stage $f \simeq 86$ MeV is the pseudoscalar decay constant in the chiral limit. 

The meson-baryon Lagrangian 
${\cal L}_{MB}^{(1)}({\cal B,U})$ at leading chiral order, ${\cal O}(p)$, involves the baryon octet $(N, \Lambda, \Sigma, \Xi)$ collected in the $3\times 3$ matrix field ${\cal B}$. The baryon fields couple to the mesonic vector and axial vector currents,
$v^\mu =  (1/2\textrm{i})[u^\dagger\partial^\mu u + u\,\partial^\mu u^\dagger]$ and $a^\mu = (1/2\textrm{i})[u^\dagger\partial^\mu u - u\,\partial^\mu u^\dagger] \equiv -(1/2)u^\mu$:
\begin{eqnarray} 
{\cal L}_{MB}^{(1)} = \mathrm{Tr}\Big(\bar{\cal B}(\textrm{i}\gamma_\mu {\cal D}^{\mu} - M_0){\cal B} - D~ \bar{\cal B}\,\gamma_\mu \gamma_5 \{a^\mu,{\cal B}\}  - F ~\bar{\cal B}\,\gamma_\mu \gamma_5 [a^\mu,{\cal B}]\Big)~~, 
\label{eq1}
\end{eqnarray}
with the chiral covariant derivative ${\cal D}^{\mu}{\cal B} = \partial^\mu {\cal B} + \textrm{i}[v^\mu,{\cal B}]$. Here $M_0$ is the baryon mass in the chiral limit. The axial vector coupling constants $D$ and $F$ are determined by neutron and hyperon beta decays.  
At next-to-leading order, ${\cal O}(p^2)$, the Lagrangian introduces several low-energy constants, $b_i$ and $d_j$:
\begin{eqnarray} 
{\cal L}_{MB}^{(2)} = b_0\, \mathrm{Tr}\big(\bar{\cal B}\,{\cal B}\big)\,\mathrm{Tr}\big(\chi_+\big)  &+& b_D \,\mathrm{Tr}\big(\bar{\cal B}\{\chi_+ , {\cal B}\}\big) + b_F\, \mathrm{Tr}\big(\bar{\cal B}[\chi_+ , {\cal B}]\big)\nonumber \\ 
&+& d_1\,\mathrm{Tr}\big(\bar{\cal B}\,\{u_\mu, [u^\mu, {\cal B}]\}\big) + d_2\,\mathrm{Tr}\big(\bar{\cal B}\,[u_\mu, [u^\mu, {\cal B}]]\big) \nonumber\\
&+& d_3\,\mathrm{Tr}\big(\bar{\cal B}\,u_\mu\big)\, \mathrm{Tr}\big({\cal B}\,u^\mu\big) + d_4\,\mathrm{Tr}\big(\bar{\cal B}\,{\cal B}\big)\,\mathrm{Tr}\big(u_\mu\, u^\mu\big)~~, 
\label{eq2}
\end{eqnarray}
where $\chi_+ =  2\,B_0\,\big(u{\cal M}u + u^\dagger {\cal M}u^\dagger\big)$ is the symmetry breaking term 
with $B_0$ representing the magnitude of the chiral condensate divided by $f^2$, and ${\cal M} = diag(m_u, m_d, m_s)$ is the quark mass matrix. At tree level in chiral perturbation theory (ChPT), the constants $b_0$, $b_D$ and $b_F$ are constrained by the baryon octet masses (their splittings and shifts from the chiral-limit baryon mass $M_0$). Note however that the present analysis goes well beyond tree level so that these constants need not be identical to the ones from ChPT. They are renormalized by loop effects taken to all orders.

\begin{figure}
\begin{center}
\includegraphics*[totalheight=2.cm]{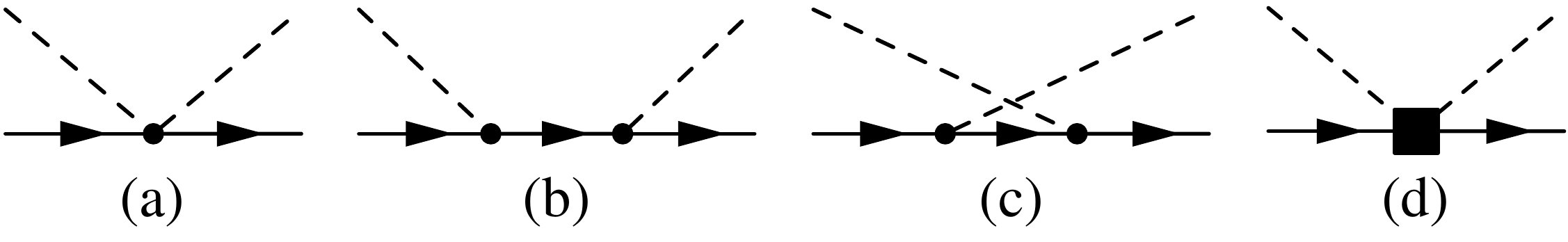}
\caption{Driving interactions generating the meson-baryon coupled-channels amplitudes: (a) leading order (Tomozawa-Weinberg) term; (b) and (c): direct and crossed Born terms; (d) NLO terms. Dashed lines represent SU(3) pseudoscalar octet mesons, solid lines refer to members of the baryon octet. }
\label{Fig1}
\end{center}
\end{figure}

We recall that ${\cal L}_{MB}^{(1)}$ of Eq. (\ref{eq1}) generates the leading Tomozawa-Weinberg (TW) terms of the interactions between the meson and baryon octets (Fig.\ref{Fig1}(a)). It also generates meson-baryon direct and crossed Born terms (Fig.\ref{Fig1}(b) and (c)) upon iteration of the pseudovector derivative coupling vertices proportional to $D$ and $F$.  The NLO terms (Fig.\ref{Fig1}(d)) derived from ${\cal L}_{MB}^{(2)}$ of Eq. (\ref{eq2}) involve, apart from the  $b_i$ coefficients, four low-energy constants $d_j$ that will be varied freely to achieve a best fit to the available $\bar{K}N$ threshold and scattering data. 

%
%

The driving meson-baryon interactions of Fig.\ref{Fig1}, derived from the NLO effective Lagrangian (\ref{eq1}) and (\ref{eq2}), serve as input interaction kernel (denoted by $\hat{V}_{ij}$) for the coupled-channels Bethe-Salpeter equations connecting meson-baryon channels $i$ and $j$. We use the full set of ten strangeness $S = -1$ channels with index assignments $i = 1, \dots, 10$, provided by the baryon and pseudoscalar meson octets and numbered in this order: $i = K^-p$, $\bar{K}^0n$,  $\pi^0\Lambda$,  $\pi^0\Sigma^0$,  $\pi^+\Sigma^-$,  $\pi^-\Sigma^+$,  $\eta\Lambda$,  $\eta\Sigma^0$,  $K^+\Xi^-$, $K^0\Xi^0$. The $\hat{V}_{ij}$ depend on the meson-baryon center-of-mass energy, $\sqrt{s}$, the scattering angles, $\Omega = \{\theta,\varphi\}$, and the baryon spin degrees of freedom, $\sigma_{i,j}$. We concentrate on s-wave driving terms, $V_{ij}(\sqrt{s}) = (1/8\pi)\sum_\sigma\int d\Omega\, \hat{V}_{ij}(\sqrt{s}, \Omega, \sigma)$, summarized and explicitly listed in Refs.~\cite{HJ2011, Borasoy2005, Borasoy2006}.
For example, the leading order Tomozawa-Weinberg term is simply\footnote{The normalization convention used here is the same as in Ref.~\cite{Borasoy2005}, with dimensionless $V_{ij}$. It differs from the one used in Ref.~\cite{HJ2011} by a factor $\sqrt{M_i M_j}$.}
\begin{eqnarray}
V_{ij}^{(TW)}(\sqrt{s}) = - {C_{ij}\over 8 f^2}\,(2\sqrt{s} - M_i - M_j)\sqrt{(M_i + E_i)(M_i + E_i)}~~,
\end{eqnarray}
where $M_i$ and $E_i = \sqrt{M_i^2 + \mathbf{q}_i^2}$ are the baryon mass and energy in channel $i$,
with $\mathbf{q}_i$ the center-of-mass momentum in that channel. The constants $C_{ij}$
are determined by SU(3) Clebsch-Gordan coefficients and given in  Refs.~\cite{Kaiser1995, Borasoy2005, Borasoy2006}.
The s-wave coupled-channels $\mathbf{T}-$matrix with elements $T_{ij}$ is found by solving the matrix integral equations
\begin{eqnarray}
\mathbf{T} = \mathbf{V} + \mathbf{V\cdot G \cdot T} = (\mathbf{V}^{-1} - \mathbf{G})^{-1}~~.
\end{eqnarray}
Here $\bf{G}$ is the Green function matrix. Its elements $G_{ij} = G_i(q^2)\,\delta_{ij}$ are the meson-baryon loop functions, 
\begin{eqnarray}
G_i(q^2) = \int {d^4k \over (2\pi)^4}{\textrm{i}\over [(q-k)^2 - M_i^2 + \textrm{i}\epsilon]
(k^2-m_i^2  + \textrm{i}\epsilon)} ~~,
\end{eqnarray}
evaluated in each channel $i$ using dimensional regularization: 
\begin{eqnarray}
G_i(\sqrt{s}) = a_i(\mu) &+&  {1\over 32\pi^2}\left[\ln\left({m_i^2\,M_i^2\over \mu^4}\right) - {M_i^2 - m_i^2\over s}\ln\left({m_i^2\over M_i^2}\right)\right] \nonumber\\
&-& {1\over 16\pi^2}\left[1 + {4|\mathbf{q}_i|\over\sqrt{s}}\,\textrm{artanh}\left({2\sqrt{s}\,|\mathbf{q}_i| \over (m_i+M_i)^2 -s}\right)\right]~~,
\label{eq:G}
\end{eqnarray}
where $m_i$ is the meson mass in channel $i$. The subtraction constants $a_i(\mu)$ act as renormalization parameters at a scale $\mu$. They cancel the scale dependent chiral logarithms and make sure that the calculated observables are scale invariant. 

\begin{figure}
\includegraphics*[totalheight=11cm]{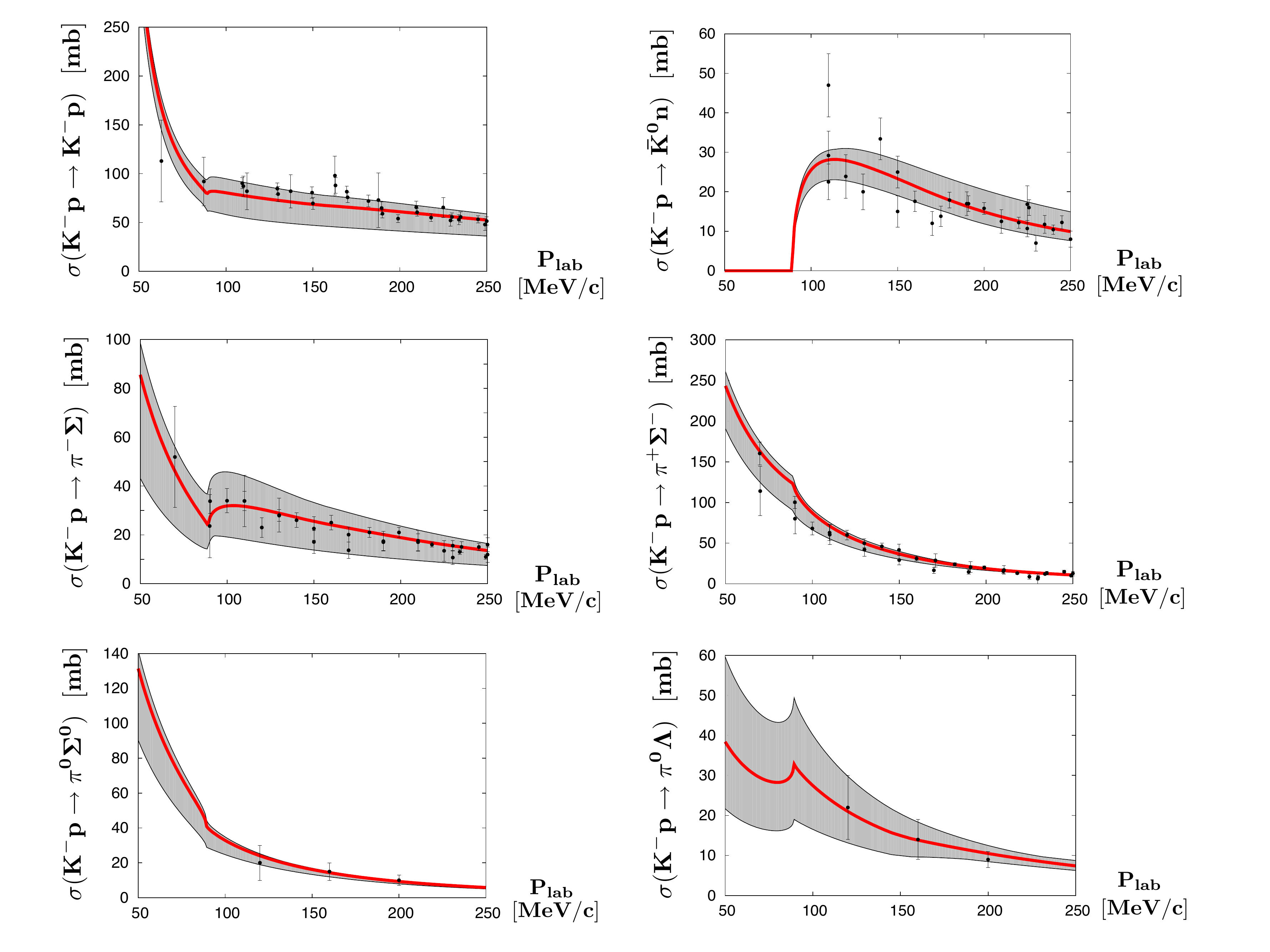}
\caption{Calculated $K^-p$ elastic, charge exchange and strangeness exchange cross sections
as function of $K^-$ laboratory momentum, compared with experimental data~\cite{cross_sections}.  
The solid curves represent best fits of the full NLO calculations to the complete data base including threshold
observables. Shaded areas give an impression of uncertainties.}
\label{Fig2}
\end{figure}

{\it Observables.} Forward scattering amplitudes and cross sections are given as 
\begin{eqnarray}
\mathbf{f}_{ij} = {1\over 8\pi\sqrt{s}}\,T_{ij}~~~~~\textrm{and} ~~~~~ \sigma_{ij}(\sqrt{s}) = {|\mathbf{q}_i|\over |\mathbf{q}_j|}{|T_{ij}(\sqrt{s})|^2\over 16\pi\,s}~~.
\end{eqnarray}
The $K^-p$ scattering length is $a(K^-p) = \lim_{th}\,{\bf f}(K^-p\rightarrow K^-p) = {\bf f}_{11}(\sqrt{s} = m_{K^-}+M_p)$. Further observables of interest are the threshold branching ratios
\begin{eqnarray}
\gamma &=& {\Gamma(K^-p\rightarrow\pi^+\Sigma^-)\over\Gamma(K^-p\rightarrow\pi^-\Sigma^+)} = {\sigma_{51}\over\sigma_{61}}~,~~R_n = {\Gamma(K^-p\rightarrow\pi^0\Lambda)\over\Gamma(K^-p\rightarrow\textrm{neutral states})} 
= {\sigma_{31}\over\sigma_{31} + \sigma_{41}}~,\nonumber\\
R_c &=& {\Gamma(K^-p\rightarrow\pi^+\Sigma^-,\,\pi^-\Sigma^+)\over\Gamma(K^-p\rightarrow\textrm{all inelastic channels})}= {\sigma_{51}+\sigma_{61}\over \sigma_{31}+\sigma_{41}+\sigma_{51}+\sigma_{61}}~~,
\end{eqnarray}
with all partial cross sections $\sigma_{ij}$ taken at $K^-p$ threshold. The (ancient but accurate) empirical values of these branching ratios \cite{BR} are listed in Table \ref{tab:results}. The energy shift and width of the 1s state of kaonic hydrogen are related to the complex $K^-p$ scattering length as
\begin{eqnarray}
\Delta E - \textrm{i}\Gamma/2 = -2\alpha^3\,\mu_r^2\,a(K^-p)\left[1+2\alpha\,\mu_r\,(1-\ln\alpha)\,a(K^-p)\right]~~,
\label{eq10}
\end{eqnarray}
with the $K^-p$ reduced mass, $\mu_r = m_K M_p/(m_K + M_p)$, and including important second order  corrections~\cite{MRR2004}. We use the accurate SIDDHARTA measurements~\cite{Bazzi2011}: 
\begin{eqnarray}
\Delta E = 283\pm36(stat)\pm6(syst)~ \textrm{eV} ~, ~~~\Gamma = 541\pm89(stat)\pm22(syst)~\textrm{eV}~~.\nonumber 
\end{eqnarray}
The available data base is completed by the collection of (less accurate) scattering cross sections~\cite{cross_sections} (see Fig.\ref{Fig2}). We do not include measured $\pi\Sigma$ mass spectra in the fitting procedure
itself but rather generate them as ``predictions" from our coupled-channels calculations.

{\it Results and discussion.}  Using the unitary coupled-channels method just described, the basic aim of the present work is to establish a much improved input set for chiral SU(3) dynamics, by systematic comparison with a variety of empirical data and with special focus on the new constraints provided by the recent kaonic hydrogen measurements~\cite{Bazzi2011}. A detailed uncertainty analysis is performed. It will be demonstrated that previous uncertainty measures~\cite{Borasoy2005,Borasoy2006} can be reduced considerably.

We have carried out $\chi^2$ fits to the empirical data set in several consecutive steps: first starting with the leading 
order (TW) terms, then adding direct and crossed Born terms, and finally using the complete NLO effective Lagrangian. The results are summarized in Table~\ref{tab:results}. All calculations have been performed using {\it empirical} meson and baryon masses.
This implies in particular that those parts of the NLO parameters $b_0, b_D$ and $b_F$ responsible for shifting the baryon octet masses from their chiral limit, $M_0$, to their physical values, are already taken care of.
The remaining renormalized parameters, denoted by  $\bar{b}_0, \bar{b}_D$ and $\bar{b}_F$, are then expected to be considerably smaller in magnitude than the ones usually quoted in tree-level chiral perturbation theory. Similar renormalization arguments imply that the pseudoscalar meson decay constants should 
be chosen at or close to their {\it physical} values~\cite{decay_constants},
\begin{eqnarray}
f_\pi = 92.4~\textrm{MeV}~, ~~~f_K = (1.19\pm0.01)\,f_\pi~,~~~f_\eta = (1.30\pm0.05)\,f_\pi~.
\label{eq:decayconst}
\end{eqnarray}
It turns out that best fit results can indeed be achieved with these physical decay constants as inputs. This is a non-trivial observation, as previous calculations have commonly used an average decay constant  as a mere fit parameter, irrespective of physical constraints.

\begin{table}[tb]
  \begin{center}
    \begin{tabular}{l|lll|l}  
               & ~~~~~~TW & ~~~~~TWB & ~~~~~NLO & ~~~~Experiment \\
      \hline
      $\Delta E$ [eV]  &~~~~~~373 & ~~~~~\,377 & ~~~~~\,306 & ~~~$283\pm 36 \pm 6$~~~~\cite{Bazzi2011} \\
      $\Gamma$ [eV]    &~~~~~~495 & ~~~~~\,514 & ~~~~~\,591 & ~~~$541\pm 89 \pm 22$~~\,\cite{Bazzi2011} \\
      $\gamma$    &~~~~ 2.36 & ~~~~~2.36 & ~~~~~2.37
      & $~~~~~2.36\pm 0.04$~~~~\cite{BR} \\
      $R_{n}$     &~~~~ 0.20 & ~~~~~0.19 & ~~~~~0.19 & $~~~0.189\pm 0.015$~~~\cite{BR} \\
      $R_{c}$    &~~~~ 0.66 & ~~~~~0.66 &~~~~~0.66 & $~~~0.664\pm 0.011$~~~\cite{BR} \\
      \hline
      $\chi^{2}$/d.o.f & ~~~~~1.12 & ~~~~~1.15 & ~~~~~0.96 &  \\
      \hline
      pole positions  ~~& ~~$1422-16i$ ~~& ~~$1421-17i$ ~~&~~$1424-26i$~~ &  \\
      ~~~~~$[\textrm{MeV}]$  &~~$1384-90i$~~ &~~$1385-105i$~~ &~~$1381-81i$~~ &  \\
    \end{tabular}
    \caption{
    Results of the systematic $\chi^{2}$ analysis using leading order (TW) plus Born terms (TWB) and full NLO schemes. Shown are the energy shift and width of the 1s state of the kaonic hydrogen ($\Delta E$ and $\Gamma$), threshold branching ratios ($\gamma$, $R_{n}$ and $R_{c}$), $\chi^{2}$/d.o.f of the fit, and the pole positions of the isospin $I=0$ amplitude in the $\bar{K}N$-$\pi\Sigma$ region.}
    \label{tab:results}
  \end{center}
\end{table}

With the TW terms alone a reasonable overall fit (with $\chi^2/\textrm{d.o.f.} = 1.12$) can be reached but the kaonic hydrogen energy shift comes out too large ($\Delta E = 373$ eV) and some of the subtraction constants $a_i$ in Eq.
(\ref{eq:G}), especially those in the $\pi\Lambda$ and $\eta\Sigma$ channels, exceed their expected ``natural" values $\sim 10^{-2}$ by more than an order of magnitude~\cite{Hyodo:2008xr}. This clearly indicates the necessity of including higher order terms in the interaction kernel $V_{ij}$. It also emphasizes the important role of the accurate kaonic hydrogen data in providing sensitive constraints.

The additional inclusion of direct and crossed meson-baryon Born terms does not change $\Delta E$ and $\chi^2/\textrm{d.o.f.}$ in any significant way. It nonetheless improves the situation considerably since the subtraction constants $a_i$ now come down to their expected ``natural" sizes.

The best fit (with $\chi^2/\textrm{d.o.f.} = 0.96$) is achieved when incorporating NLO terms in the calculations. The 
inputs used are: the decay constants  $f_\pi = 92.4$ MeV, $f_K = 110.0$ MeV, $f_\eta = 118.8$ MeV, 
and axial vector couplings $D = 0.80, F= 0.46$ (i.e. $g_A = D+F = 1.26$); subtraction constants at a renormalization scale $\mu = 1$ GeV (all in units of $10^{-3}$): $a_1 = a_2 = -2.38, ~a_3 = -16.57, ~a_4 = a_5 = a_6 = 4.35,  ~a_7 = -0.01, ~a_8 = 1.90, ~a_9 = a_{10} = 15.83$; and NLO parameters (in units of $10^{-1}$ GeV$^{-1}$):  $\bar{b}_0 = -0.48,~\bar{b}_D = 0.05,~\bar{b}_F = 0.40,~d_1= 0.86,~d_2 = -1.06,~d_3 = 0.92,~
d_4 = 0.64$.
Within the set of altogether ``natural"-sized constants $a_i$ the relative importance of the $K \Xi$ channels involving double-strangeness exchange is worth mentioning.

As seen in Table~\ref{tab:results}, the results are in excellent agreement with threshold data. The same input reproduces the whole set of $K^-p$ cross section measurements as shown in Fig. \ref{Fig2} (Coulomb interaction effects are included in the diagonal $K^{-}p\to K^{-}p$ channel as in Ref.~\cite{Borasoy2005}).  A systematic uncertainty analysis has been performed by varying the input parameters within the range permitted by the uncertainty measures of the experimental data. A detailed description of this analysis will be given in a longer forthcoming paper~\cite{IHW2011}. 

%
%

\begin{figure}[htb]
\begin{minipage}[t]{7cm}
\includegraphics[width=6cm]{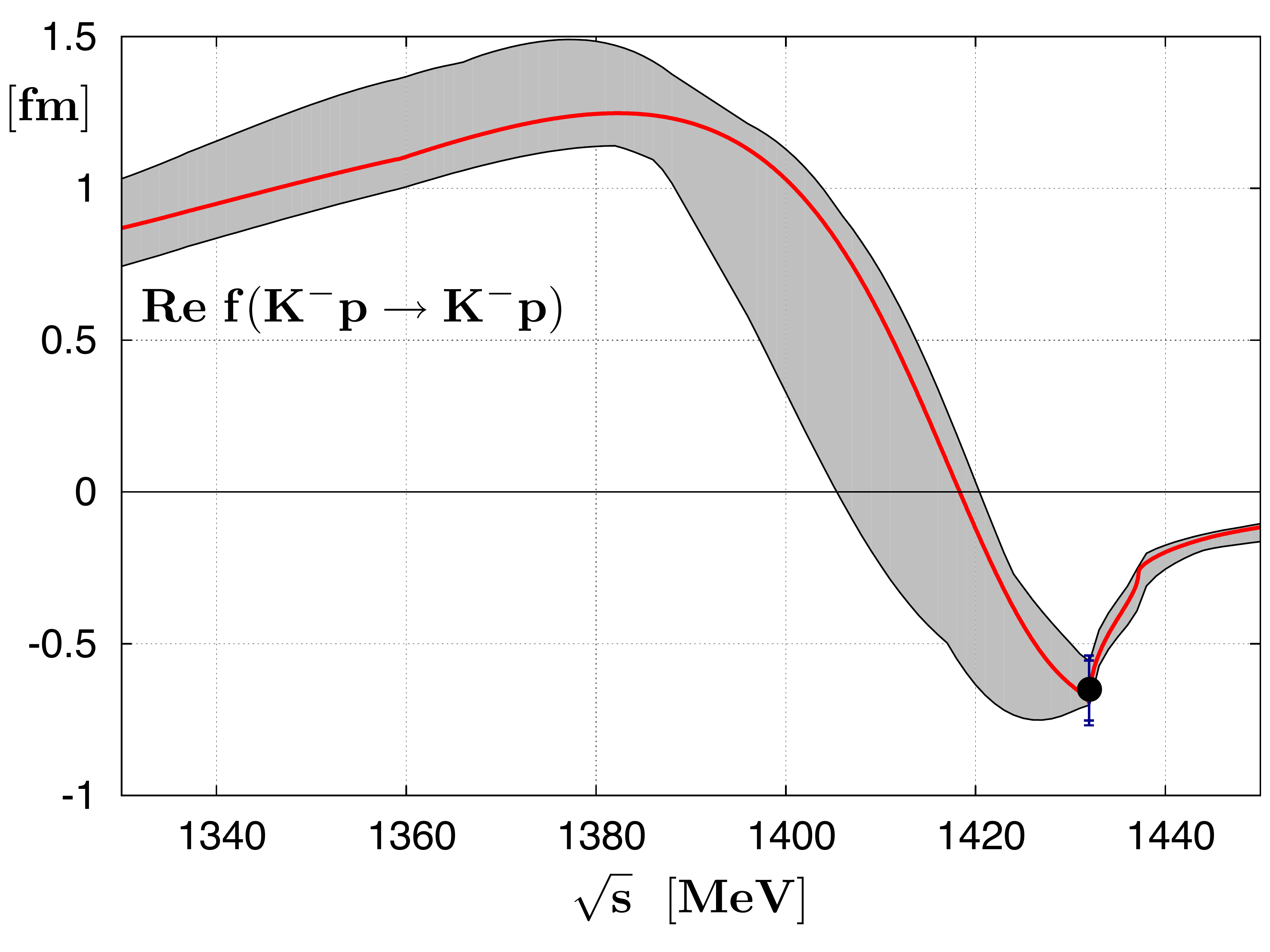}
\end{minipage}
\hspace{\fill}
\begin{minipage}[t]{7cm}
\includegraphics[width=6.2cm]{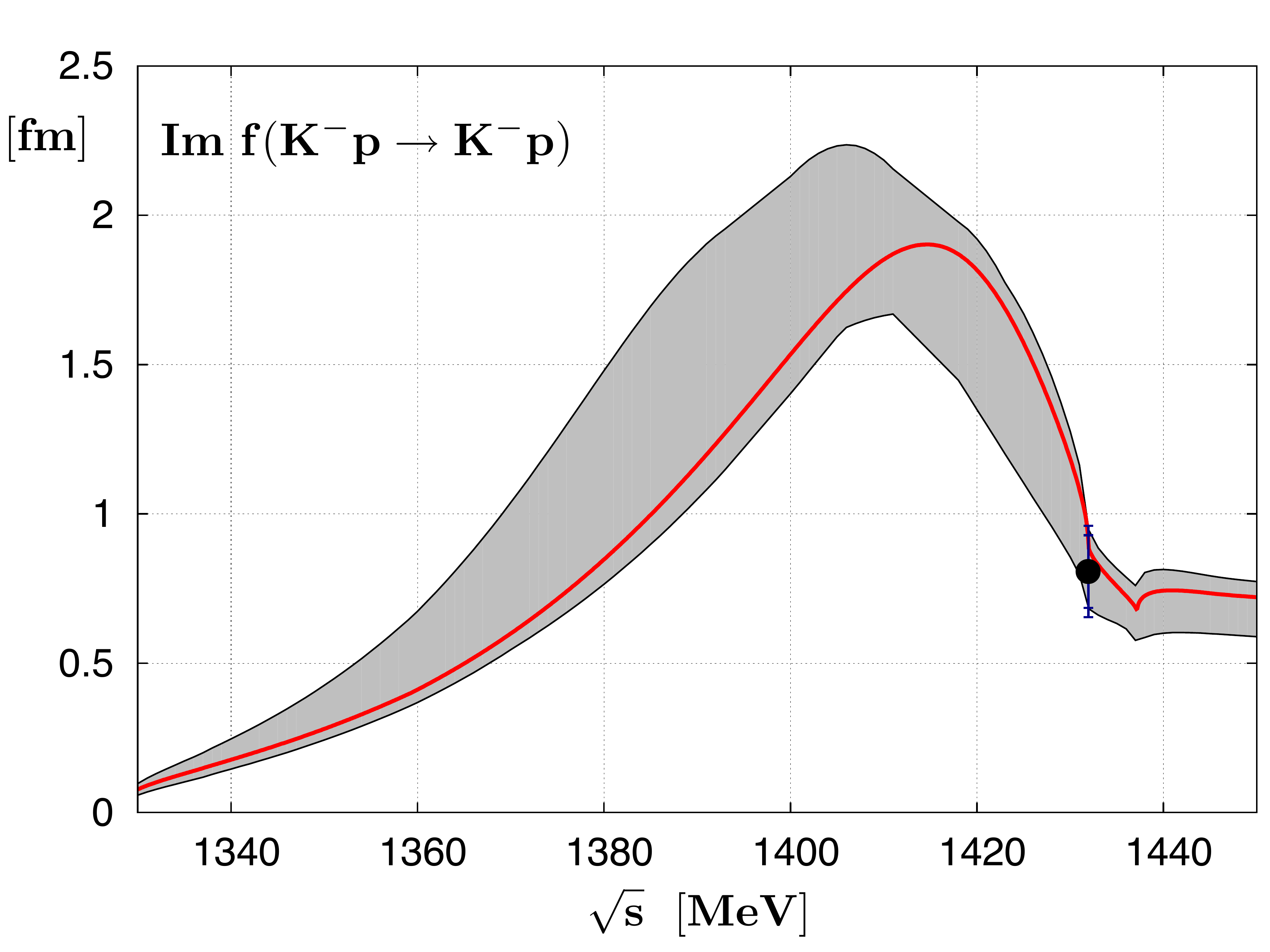}
\end{minipage}
\caption{Real part (left) and imaginary part (right) of the $K^-p \rightarrow K^-p$ forward scattering amplitude
extrapolated to the subthreshold region. The empirical real and imaginary parts of the $K^-p$ scattering length deduced from the recent kaonic hydrogen measurement (SIDDHARTA~\cite{Bazzi2011}) are indicated by the dots including statistical and systematic errors. The shaded uncertainty 
bands are explained in the text.} 
\label{Fig3}
\end{figure}

Equipped with the best fit to the observables at $K^-p$ threshold and above, an optimized prediction for the 
subthreshold extrapolation of the complex s-wave $K^-p\rightarrow K^-p$ amplitude can now be given.
The result is shown in Fig. \ref{Fig3}, including again a conservative uncertainty estimate. The real and 
imaginary parts of this amplitude display as expected the $\Lambda(1405)$ resonance as a quasibound
$\bar{K}N$ $(I=0)$ state embedded in the $\pi\Sigma$ continuum. The present NLO calculation confirms
the two-poles scenario~\cite{Jido2003,HW2008} of the coupled $K^-p\leftrightarrow \pi\Sigma$ system.
Using the best-fit input, the resulting locations of the two poles in the complex energy plane are as follows:
``upper" pole ($\bar{K}N$-dominated): 1424 - i 26 MeV; ``lower" pole ($\pi\Sigma$-dominated):
1381 - i 81 MeV. Unlike previously found patterns in which the location of the lower pole has been 
subject to large model uncertainties, the pole positions now remain remarkably stable with respect to changes
of the input. The shift of the real parts of both these pole positions from the ``TW" and ``TW + Born terms" steps 
to the full NLO calculation is less than 5 MeV. The corresponding change in the imaginary parts
is only slightly larger (between about 10 and 20 MeV).

%
%

The $K^-p$ scattering length, $a(K^-p)$, deduced from the
kaonic hydrogen measurements~\cite{Bazzi2011} and with inclusion of Coulomb 
corrections (see Eq.(\ref{eq10})) is:
\begin{eqnarray}
\textrm{Re}\,a(K^-p) = -0.65 \pm 0.10 ~~\textrm{fm}~~,~~~\textrm{Im}\,a(K^-p) = 0.81\pm 0.15 ~~\textrm{fm}~~,
\label{scattlength}
\end{eqnarray}
with an error estimate based on the uncertainties assigned to the measured kaonic hydrogen
energy shift and width. Our best fit NLO result, $a(K^-p) = -0.70 + \textrm{i}\, 0.89$ fm, is perfectly consistent with 
Eq.\,(\ref{scattlength}). Note that this new determination of the $K^-p$ scattering length has shifted
quite significantly in the value of Re$\,a(K^-p)$ from previous ones~\cite{Borasoy2005,Borasoy2006,ORB2002,
Hyodo2003}, mainly because of the new constraints from the much improved SIDDHARTA data.

{\it Summary.} Given the significantly more accurate constraints from the new kaonic hydrogen
measurements, an improved theory of low-energy antikaon-nucleon interactions on the basis of chiral SU(3)
effective field theory with coupled-channels is now at hand. The results and conclusions are summarized as follows:

1) Kaonic hydrogen data are now consistent with low-energy $K^-p$ elastic, charge exchange and strangeness exchange cross sections.

2) The present next-to-leading-order (NLO) analysis indicates a well convergent hierarchy of driving terms for
coupled-channels dynamics derived from the chiral SU(3) meson-baryon effective Lagrangian. The 
Tomozawa-Weinberg terms dominate, Born terms are significant, while a ``best fit" nonetheless requires NLO contributions, though with relatively small coefficients.

3) A new, more accurate determination of the $K^-p$ scattering length has been presented.

4) As an important result of the present analysis, the best NLO fit systematically prefers {\it physical}
values of the decay constants $f_\pi, f_K$ and $f_\eta$. This is actually a (successful) test of consistency,
given that physical meson and baryon masses and empirical baryon axial vector coupling constants
are used as input.

5) The two-poles scenario of $\bar{K}N$ and $\pi\Sigma$ coupled-channels dynamics is re-confirmed.
The predictions for the pole positions in the complex energy plane have been sharpened.

6) Uncertainties in the subthreshold extrapolation of the $\bar{K}N$ amplitude are reduced as compared to 
previous work.

A more detailed presentation including further results and predictions, e.g. on $\pi\Sigma$ invariant mass
spectra and the $K^-n$ scattering length, is in preparation. \\

{\it Acknowledgements.} 
We thank Avraham Gal for helpful comments and discussions. This work has been performed under the joint research cooperation agreement between RIKEN and Technische Universit\"at M\"unchen. It
is partly supported by BMBF, GSI, the DFG Cluster of Excellence ``Origin and Structure of the Universe", 
and by the Grant-in-Aid for Scientific Research from MEXT and JSPS (No.\ 21840026 and 23-8687).
T.H. thanks for support from the Global Center of Excellence Program by MEXT, Japan through the Nanoscience and Quantum Physics Project of the Tokyo Institute of Technology.

\end{document}